\documentclass[12pt]{article}
\usepackage{graphicx}
\begin{document}

\vbox{\vspace{5ex}}

\begin{center}
{\Large \bf Time separation as a hidden variable to the Copenhagen
school of quantum mechanics}

\vspace{2ex}
Y. S. Kim \footnote{electronic address: yskim@physics.umd.edu}\\
Center for Fundamental Physics, University of Maryland,\\
College Park, Maryland 20742, U.S.A.\\

\vspace{2ex}

Marilyn E. Noz \footnote{electronic address: nozm01@med.nyu.edu}\\
Department of Radiology, New York University,\\
New York, New York 10016, U.S.A.\\

\end{center}
\vspace{3ex}

\begin{abstract}
The Bohr radius is a space-like separation between the proton and
electron in the hydrogen atom.  According to the Copenhagen school
of quantum mechanics, the proton is sitting in the absolute Lorentz
frame.  If this hydrogen atom is observed from a different Lorentz
frame, there is a time-like separation linearly mixed with the Bohr
radius.  Indeed, the time-separation is one of the essential variables
in high-energy hadronic physics where the hadron is a bound state of
the quarks, while thoroughly hidden in the present form of quantum
mechanics.  It will be concluded that this variable is hidden in
Feynman's rest of the universe.  It is noted first that Feynman's
Lorentz-invariant differential equation for the bound-state quarks
has a set of solutions which describe all essential features of
hadronic physics. These solutions explicitly depend on the time
separation between the quarks.  This set also forms the mathematical
basis for two-mode squeezed states in quantum optics, where both
photons are observable, but one of them can be treated a variable
hidden in the rest of the universe.  The physics of this two-mode
state can then be translated into the time-separation variable in
the quark model.  As in the case of the un-observed photon, the hidden
time-separation variable manifests itself as an increase in entropy
and uncertainty.
\end{abstract}

\section{Introduction}\label{intro}
The Bohr radius is a very important parameter in the present form of
quantum mechanics.  Niels Bohr spent much of his research life on
the hydrogen atom.  He also had a great respect for Einstein.
Whenever he mentions ``space''  he also adds ``time.''  Yet, he never
thought about the proton in other than the absolute frame.  Nor did
Einstein raise this issue.  At their time, it was beyond their
imagination that bound-state particles could move with relativistic
speed.
\par
\begin{figure}[thb]
\centerline{\includegraphics[scale=0.3]{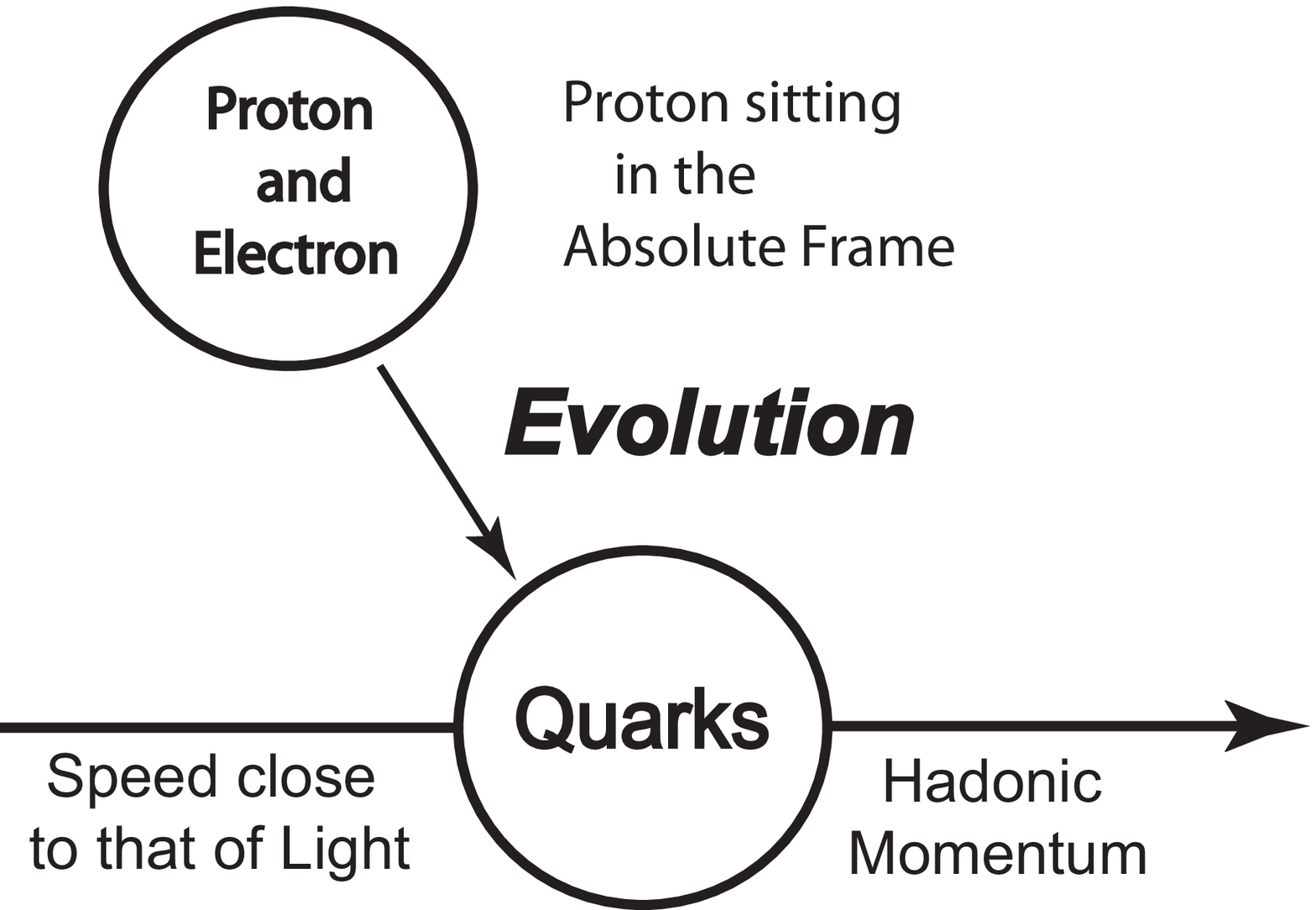}
\hspace{10mm}
\includegraphics[scale=0.3]{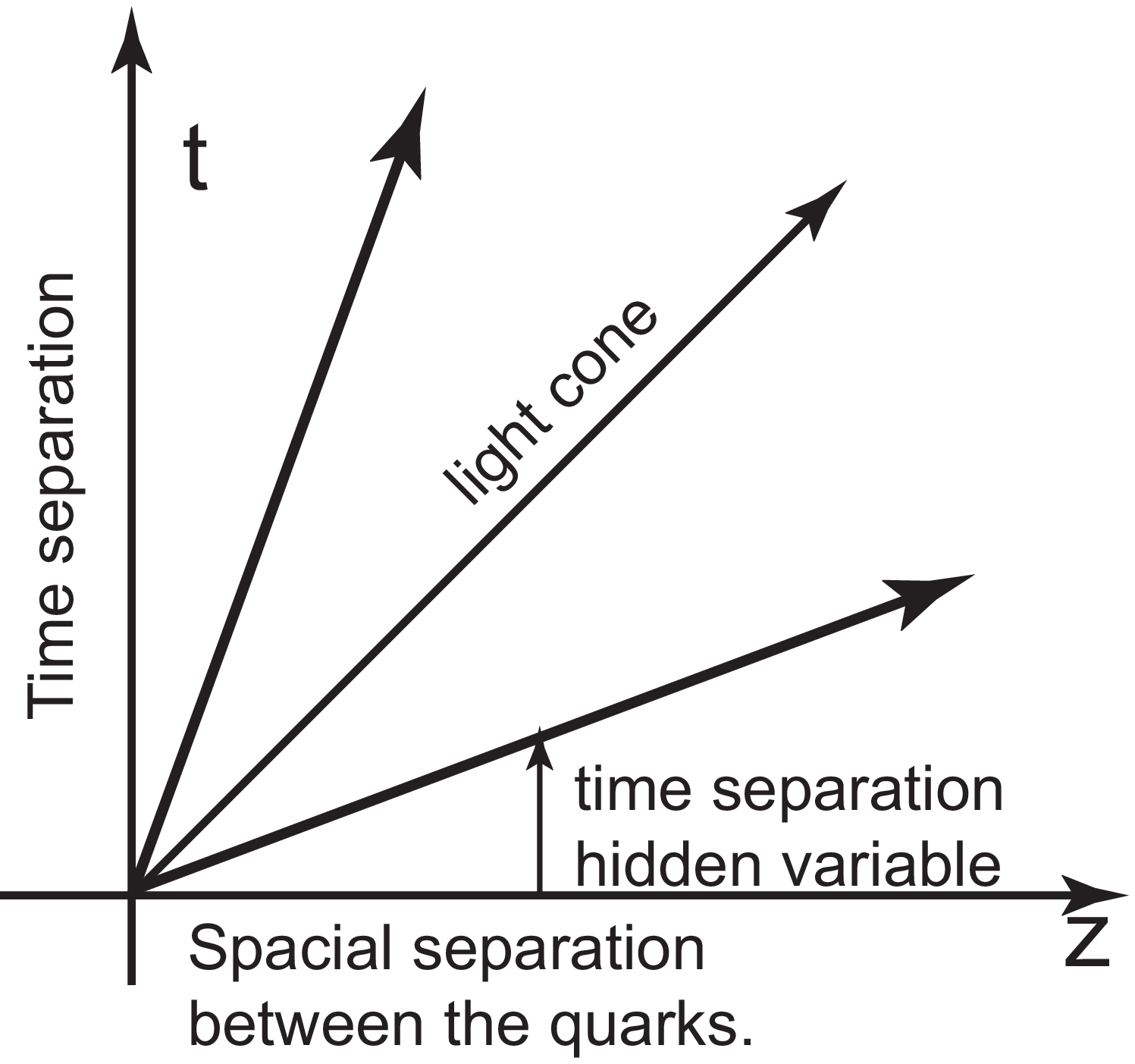}}
\caption{Evolution of the hydrogen atom.  The Bohr radius measures the
spacial separation between the proton and electron in the hydrogen
atom, without time separation.  Let us assume that this separation is
zero in the frame where the hydrogen atom is at rest.  Then the
time-separation becomes prominent to a moving observer.}\label{evol}
\end{figure}

There are now composite particles moving with speed close to that of
velocity of light, but they are not hydrogen atoms.  They are protons
coming out from particle accelerators, and each proton is a bound
state of quarks.  As far as quantum bound states are concerned, the
hydrogen atom went through an evolutionary process as described in
Fig.~\ref{evol}.
\par
In this paper, we start with the Lorentz-invariant oscillator equation
of Feynman {\em et al.}~\cite{fkr71}, and note that there is a
Lorentz-covariant set of solutions representing Wigner's little group
dictating the internal space-time symmetry of particles~\cite{wig39}.
This set also describes the essential feature of high-energy hadronic
physics~\cite{knp86}.   Furthermore, these covariant solutions carry
quantum probability interpretation~\cite{knp86}.
\par
It is noted also that this set constitutes the mathematical basis
for two-photon coherent states or two-mode squeezed state, where
both photons are observable~\cite{knp91}.  This two-photon system
allows us to consider when one of the photons is not
observed~\cite{ekn89}.
\par
The Lorentz-covariant oscillator formalism thus allows us to study
what happens when the time-separation variable is not observed.
We thus conclude that it is hidden in Feynman's rest of the
universe~\cite{fey72,hkn99ajp}, which is well defined in terms of
the two-mode squeezed state.  Since this variable is hidden, it
causes an increase in entropy and uncertainty.
\par
In Sec.~\ref{feyninv}, we write down the Lorentz-invariant
differential equation which Feynman {\em et al.} used to study the
hadronic mass spectra~\cite{fkr71}.  In Sec.~\ref{wiglittle},
it is shown that the Feynman equation has a set of solutions
that can represent Wigner's $O(3)$-like little group for massive
particles.  In Sec.~\ref{diracqm}, it is shown that this set of
oscillator solutions can combine Dirac's efforts to combine quantum
mechanics and relativity~\cite{dir27,dir45,dir49}.  It is shown that
the same set of solutions can be obtained from the system of two
coupled harmonic oscillators which forms the mathematical basis for
two-mode squeezed states in quantum optics~\cite{knp91}.  Finally,
in Sec.~\ref{restof}, we illustrate Feynman's rest of universe using
the coupled harmonic oscillators and two-mode squeezed
states~\cite{fey72,hkn99ajp}.  We then show that the time-separation
variable can be interpreted as one of the oscillator variables not
observed.  The result is an increase in statistical entropy and
uncertainty, as is expected from hidden variables.

\section{Feynman's Lorentz-invariant Equation}\label{feyninv}
For solving practical problems in quantum mechanics, we use
the Schr\"odinger wave equation.   For scattering problems, we use
running-wave solutions.  For bound states, we obtain standing-wave
solutions with their boundary conditions.  Indeed, the localization
boundary condition leads to discrete energy levels.
\par
For scattering problems, we now have Lorentz-covariant quantum
field theory with its scattering matrix formalism and Feynman diagrams.
For bound state problems, there had been attempts in the past to
understand bound-state problems using the S-matrix method.  However,
it was noted that the bound-state poles of the S-matrix do not always
gurantee the localization of wave functions~\cite{kim65,gfchew67}.
\par
In 1971, Feynman {\em et al.}~\cite{fkr71} published a paper
containing the Lorentz-invariant differential equation
\begin{equation}\label{diff11}
\left\{-\frac{1}{2}\left[\left(\frac{\partial}
                               {\partial x^a_{\mu}}\right)^2 +
\left(\frac{\partial}{\partial x^b_{\mu}}\right)^2 \right]  +
\frac{1}{16}\left(x^a_{\mu} -x^b_{\mu}\right)^2 + m_0^2\right\}
\phi\left(x^a_{\mu}, x^b_{\mu}\right) = 0 ,
\end{equation}
for a hadron consisting of two quarks bound-together a harmonic
oscillator potential.  The space-time quark coordinates are
$x^a_{\mu}$ and $x^b_{\mu}$.  They then wrote down the hadronic
and the quark separation coordinates as
\begin{equation}\label{coord}
 X_\mu = \frac{1}{2}\left(x^a_{\mu} + x^b_{\mu} \right), \qquad
 x_\mu = \frac{1}{2\sqrt{2}}\left(x^a_{\mu} - x^b_{\mu}\right) ,
\end{equation}
respectively.  We can now consider the solution of the  form
\begin{equation}
\phi\left(x^a_{\mu},x^b_{\mu}\right) = f\left(X_{\mu}\right)
    \psi\left(x_{\mu}\right),
\end{equation}
where $f\left(X_mu\right)$ and $\psi\left(x_\mu\right)$ are for
a free hadron and for the quarks inside the hadron respectively.
$f\left(X_{\mu}\right)$ satisfies the Klein-Gordeon equation,
and takes the form
\begin{equation}
f(X) = \exp{ \left( \pm iP\cdot X \right) } ,
\end{equation}
with
\begin{equation}
 - P^2 = m_0^2 + (\lambda + 1) .
 \end{equation}
The quark wave function satisfies the differential equation
\begin{equation}\label{diff33}
\frac{1}{2} \left\{-\left(\frac{\partial}{\partial x_{\mu}}\right)^2
 + x_{\mu}^2 \right\} \psi\left(x_{\mu}\right) = (\lambda + 1)
 \psi \left(x_{\mu}\right) .
\end{equation}
This differential equation of Eq.(\ref{diff33}) is a
Lorentz-invariant equation, but its solution can take different
forms depending on the separable coordinate systems with their
boundary conditions.  The problem is to choose the set of solutions
which can tell us physics properly based on the existing rules of
quantum mechanics and relativity.
\par
If we ignore the time-like variable Eq.(\ref{diff33}), it becomes the
Schr\"odinger-type equation for a harmonic oscillator.  The problem
is the existence of the time-like variable.  If we ignore it, it is
the equation for non-relativistic quantum mechanics, but it is not
Lorentz-invariant.  If we include it, we have to give a physical
interpretation to this variable.

\section{Solutions representing Wigner's Little Group}\label{wiglittle}
In 1979, Kim, Noz, and Oh published a paper on representations of
the Poincar\'e group using a set of solutions of the oscillator
equation of Eq.(\ref{diff33})~\cite{kno79jmp}.  Later in 1986, Kim
and Noz in their book~\cite{knp86} noted that this set corresponds
to a representation of Wigner's $O(3)$-like little group for massive
particles.  If a particle has a non-zero mass, there is a Lorentz
frame in which the particle is at rest.  Wigner's little group
then becomes that of group $O(3)$ which is the three-dimensional
rotation group~\cite{wig39}.
\par
The solution of the Lorentz-invariant equation contains both
space-like and time-like wave functions, but we can keep the
time-like component to its ground state, in accordance with
Dirac's c-number time-energy uncertainty relation~\cite{dir27}.
The wave function still retains the $O(3)$-like symmetry.  The
solution takes the form
\begin{equation}\label{sol00}
\psi(x,y,z,t) =
 \left[\left(\frac{1}{\pi}\right)^{1/4}
    \exp{\left(\frac{-t^2}{2}\right)}\right]\psi(x,y,x) .
\end{equation}
As for the spatial part of the differential equation, we note
that it is the equation for the three-dimensional oscillator.
We can solve this equation with both the Cartesian and
spherical coordinates.  If we use the spherical system with
$(r, \theta, \phi)$ as the variables, the solution should take
the form
\begin{equation}\label{sol22}
\psi(x,y,z) =  R_{\lambda,\ell}(r)Y_{\ell,m}(\theta,\phi)
\exp{\left\{ -\left(\frac{x^2 + y^2 +z^2}{2}\right)\right\}} ,
\end{equation}
where $Y_{\ell,m}(\theta,\phi)$ is the spherical harmonics, and
$R_{\lambda,\ell}(r)$ is the normalized radial wave function
with $r = \sqrt{x^2 + y^2 + z^2}.$  The $\lambda$ and $\ell$
parameters specify the mass and the internal spin of the hadron
respectively, as required by Wigner's representation
theory~\cite{wig39,knp86}.
\par
This oscillator wave function is separable also in the Cartesian
coordinates, and the solution can be written
\begin{equation}\label{sol33}
 \psi(x,y,z)=
 \left[\frac{1}{\pi\sqrt{\pi} 2^{(a+b+n)}a!b!n!}\right]^{1/2}
          H_a(x) H_b(y) H_n(z)
 \exp{\left\{-\left(\frac{x^2 + y^2 +z^2}{2}\right)\right\}} ,
\end{equation}
where $H_n(z)$ is the Hermite polynomial, and $\lambda$ of
Eq.(\ref{diff33}) is $(a + b + n).$
\par
When we boost this solution along the $z$ direction, the
Cartesian form of Eq.(\ref{sol33}) is more convenient.  Since
the transverse $x$ and $y$ coordinates are not affected by this
boost, we can separate out these variables in the oscillator
differential equation of Eq.(\ref{diff33}), and consider the
differential equation
\begin{equation}\label{diff44}
\frac{1}{2} \left\{\left[-\left(\frac{\partial}{\partial z}\right)^2
 + z^2 \right]
 -\left[-\left(\frac{\partial}{\partial t}\right)^2 +
 t^2\right]\right\}\psi(z,t) = n \psi(z,t) .
\end{equation}
This differential equation remains invariant under the Lorentz boost
\begin{equation}\label{boostm}
z \rightarrow (\cosh\eta)z + (\sinh\eta)t , \qquad
 t \rightarrow (\sinh\eta)z + (\cosh\eta)t .
\end{equation}
In terms of the hadronic velocity $v$, $e^{\eta}$ takes the form
\begin{equation}
e^{\eta} = \sqrt{\frac{1 + v/c}{1 - v/c}} .
\end{equation}
\par
If we suppress the excitations along the $t$ coordinate, the
normalized solution of this differential equation is
\begin{equation}\label{sol44}
\psi(z,t) =  \left(\frac{1}{\pi 2^{n}n!} \right)^{1/2}
 H_n(z)\exp{\left\{-\left(\frac{z^2 +t^2}{2}\right)\right\}} .
\end{equation}
If we boost the hadron along the $z$ direction, the coordinate
variables $z$ and $t$ should be replaced respectively by $z'$
and $t'$ of Eq.(\ref{boostm}), and the wave function becomes
uncontrollable.
\par
\section{Dirac's attempts to combine quantum mechanics and special
relativity}\label{diracqm}
Paul A. M. Dirac published a number of important papers on combining
quantum mechanics with relativity.  In 1927~\cite{dir27}, Dirac
noted that there is a time-energy uncertainty relation without
time-like excitations.  He pointed out that this space-time
asymmetry causes a difficulty in combining quantum mechanics with
special relativity.
\par
In 1945~\cite{dir45}, Dirac constructed four-dimensional
harmonic oscillator wave functions including the time variable.
His oscillator wave functions took normalizable Gaussian form,
but he did not attempt to give a physical interpretation to this
mathematical device.
\par
It is remarkable that the oscillator representation given in
Sec.~\ref{wiglittle} addresses Dirac's concerns in all of
his papers mentioned above.
In his 1949 paper~\cite{dir49}, Dirac introduced his light-cone
variables defined as
\begin{equation}\label{lcvari}
u = \frac{z + t}{\sqrt{2}} , \qquad v = \frac{z - t}{\sqrt{2}} .
\end{equation}
Then the boost transformation of Eq.(\ref{boostm}) takes the form
\begin{equation}\label{lorensq}
u \rightarrow e^{\eta } u , \qquad v \rightarrow e^{-\eta } v .
\end{equation}
The $u$ variable becomes expanded while the $v$ variable becomes
contracted.  Their product
\begin{equation}
uv = {1 \over 2}(z + t)(z - t) = {1 \over 2}\left(z^2 - t^2\right)
\end{equation}
remains invariant.  Indeed, in Dirac's picture, the Lorentz boost is a
squeeze transformation.
\par
\begin{figure}
\centerline{\includegraphics[scale=0.5]{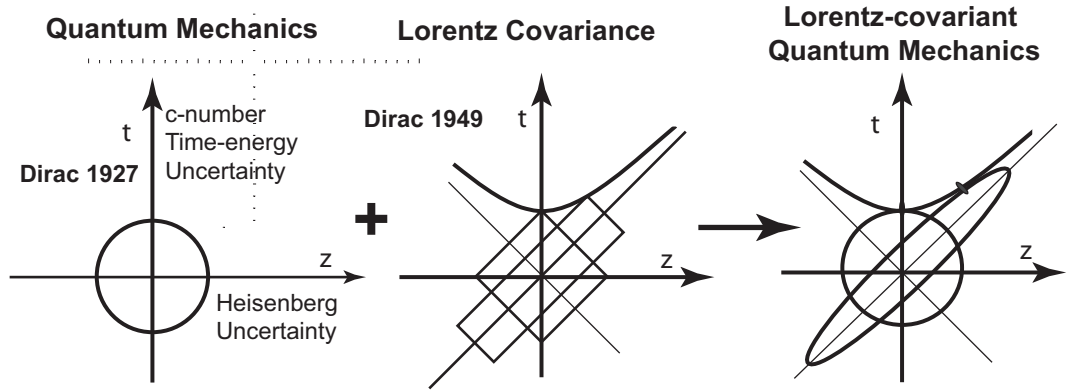}}
\caption{Lorentz-squeezed hadrons.  Feynman's proposal leads us to combine
Dirac's quantum mechanics and his light-cone representation of Lorentz
boost to generate Lorentz-squeezed hadrons.}\label{diracqm55}
\end{figure}
\par
In this new notation, the wave function of Eq.(\ref{sol44}) takes
the form
\begin{equation}\label{cwf22}
 \psi_{\eta}^{n}(x,t) = \left[1 \over \pi n! 2^{n} \right]^{1/2}
      H_{n}\left({e^{-\eta}u +  e^{\eta} v \over \sqrt{2}}\right)
  \exp{\left\{-\left(\frac{e^{-2\eta}u^{2} + e^{2\eta}v^{2}}{2}
    \right)\right\}} ,
\end{equation}
for the moving hadron.  The Gaussian factor in this expression
determines the space-time localization property of all excited-state
wave functions.  We can now combine Dirac's 1927~\cite{dir27},
1945~\cite{dir45}, and 1949~\cite{dir49} papers into
Fig.~\ref{diracqm55}.
\par
This squeeze property has been experimentally verified in various
observations in high-energy physics, including Feynman's parton
picture~\cite{fey69a,kn77par,kim89}.
\par
In 1963~\cite{dir63}, Dirac used two coupled oscillators to
understand Lorentz transformations.  Following the spirit of
Dirac, let us start with two independent oscillators.  The
wave function for this system is
\begin{equation}
 \psi\left(x_{1},x_{2}\right) = \chi_{n_1}\left(x_{1}\right)
      \chi_{n_2}\left(x_{2}\right) ,
\end{equation}
where $\chi_n(z)$ is the $n$-th excited-state oscillator wave function
which takes the form
\begin{equation}\label{chi11}
    \chi_n (x) = \left[\frac{1}{\sqrt{\pi}2^n n!}\right]^{1/2}
             H_n(x) \exp{\left(\frac{-x^2}{2}\right)} .
\end{equation}
If the $x_2$ coordinate is in its ground state, the wave function
becomes
\begin{equation}\label{cwf88}
 \psi\left(x_{1},x_{2}\right)
  = \chi_n\left(x_1\right) \chi_0\left(x_2\right)
  = \left[\frac{1}{\pi 2^n n!}\right]^{1/2} H_{n} \left(x_1\right)
    \exp{\left[- \frac{1}{2}\left(x_1^2 + x_2^2 \right)\right]} ,
\end{equation}
with $n = n_1$.  In order to couple these two oscillators, we
introduce the normal coordinates
\begin{equation}\label{normal11}
 y_1 = \frac{1}{\sqrt{2}} \left( x_{1} + x_{2}\right), \qquad
 y_2 = \frac{1}{\sqrt{2}} \left( x_{1} - x_{2}\right) ,
\end{equation}
and squeeze these variables:
\begin{equation}
 y_{1} \rightarrow e^{\eta} y_1,  \qquad
 y_{2} \rightarrow e^{-\eta} y_2,
\end{equation}
Then the squeezed wave function takes the form of Eq.(\ref{cwf22})
with $u$ and $v$ replaced by $y_1$ and $y_2$ respectively, or
with $z$ and $t$ by $x_1$ and $x_2$ respectively.  Furthermore,
this wave function can be expanded in terms of $\chi_n(x)$~\cite{knp91}.
If $n = 0$ in Eq.(\ref{cwf88}), the wave function becoms Gaussian,
and its squeezed form becomes
\begin{equation}\label{cwf55}
   \psi_{\eta}^{0}\left(x_1,x_2\right) =
   \left(\frac{1}{\cosh\eta}\right)
     \sum_{k} (\tanh\eta)^{k}\chi_{k}
     \left(x_1\right)\chi_{k}\left(x_2\right) .
\end{equation}
This expression is for the two-mode squeezed states in quantum
optics~\cite{knp91,dir63}, where $\chi\left(x_1\right)$ and
$\chi\left(x_2\right)$ are the states of the first and second
photons respectively.  Using this formula, it is possible to study
what happens if the second photon is not observed~\cite{ekn89}.
\par

\section{Hidden in Feynman's Rest of the Universe}\label{restof}
Throughout this paper, the time-separation variable played a major
role in the covariant formulation of the harmonic oscillator wave
functions.  It should exist wherever the space separation exists.
The Bohr radius is the measure of the separation between the proton
and electron in the hydrogen atom.  If this atom moves, the radius
picks up the time separation, according to Einstein~\cite{kn06aip}.
\par
On the other hand, the present form of quantum mechanics does not
include this time-separation variable.  The best way we can do at
the present time is to treat this time-separation as a variable in
Feynman's rest of the universe~\cite{hkn99ajp}.  In his book on
statistical mechanics~\cite{fey72}, Feynman states
\begin{quote}
{\it When we solve a quantum-mechanical problem, what we really do
is divide the universe into two parts - the system in which we are
interested and the rest of the universe.  We then usually act as if
the system in which we are interested comprised the entire universe.
To motivate the use of density matrices, let us see what happens
when we include the part of the universe outside the system.}
\end{quote}
This abstract statement can be studied in terms of two coupled
oscillators~\cite{hkn99ajp}, and also in terms of two-mode squeezed
states~\cite{ekn89}.
The failure to include what happens outside the system results in
an increase of entropy.  The entropy is a measure of our ignorance
and is computed from the density matrix.  The density matrix is
needed when the experimental procedure does not analyze all relevant
variables to the maximum extent consistent with quantum mechanics.
If we do not take into account the time-separation variable, the
result is therefore an increase in entropy~\cite{kiwi90pl,kim07}.
\par
From the covariant oscillator wave functions defined in this section,
the pure-state density matrix is
\begin{equation}\label{den11}
 \rho_\eta^{n}(z,t;z',t') = \psi_\eta^{n}(z,t) \psi_\eta^{n} (z',t') ,
\end{equation}
which satisfies the condition $\rho^2  = \rho: $
\begin{equation}
 \rho_\eta^{n}(z,t;x',t') = \int \rho_\eta^{n}(z,t;x",t")
  \rho_\eta^{n}(z",t";z',t') dz" dt" .
\end{equation}
In order to simplify our discussion without sacrificing physics, we
carry out our calculation for the ground state only with $n = 0$.
The computation can be extended for excited states.
\par
\begin{figure}
\centerline{\includegraphics[scale=0.29]{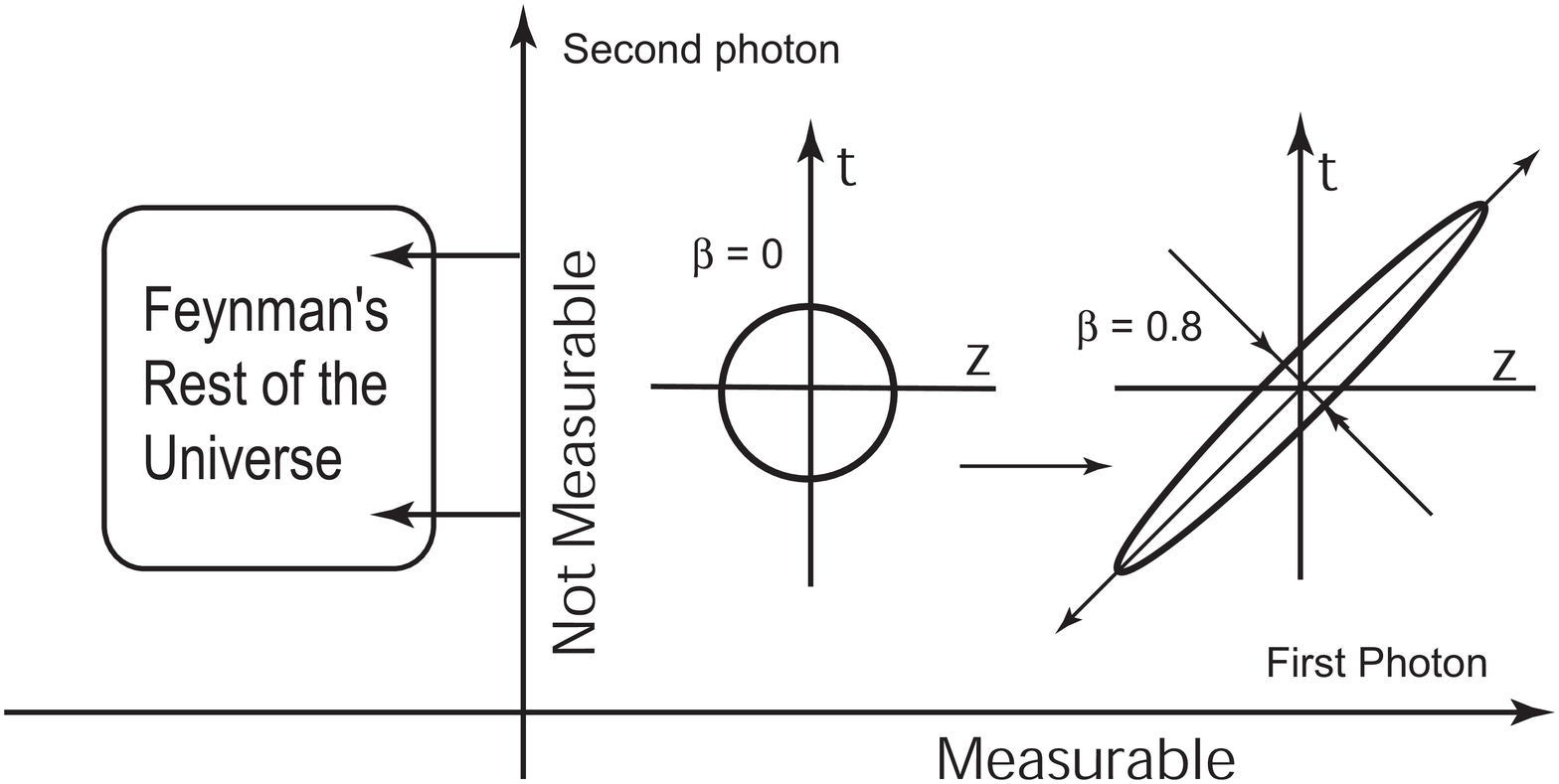}
\hspace{3mm}
\includegraphics[scale=0.24]{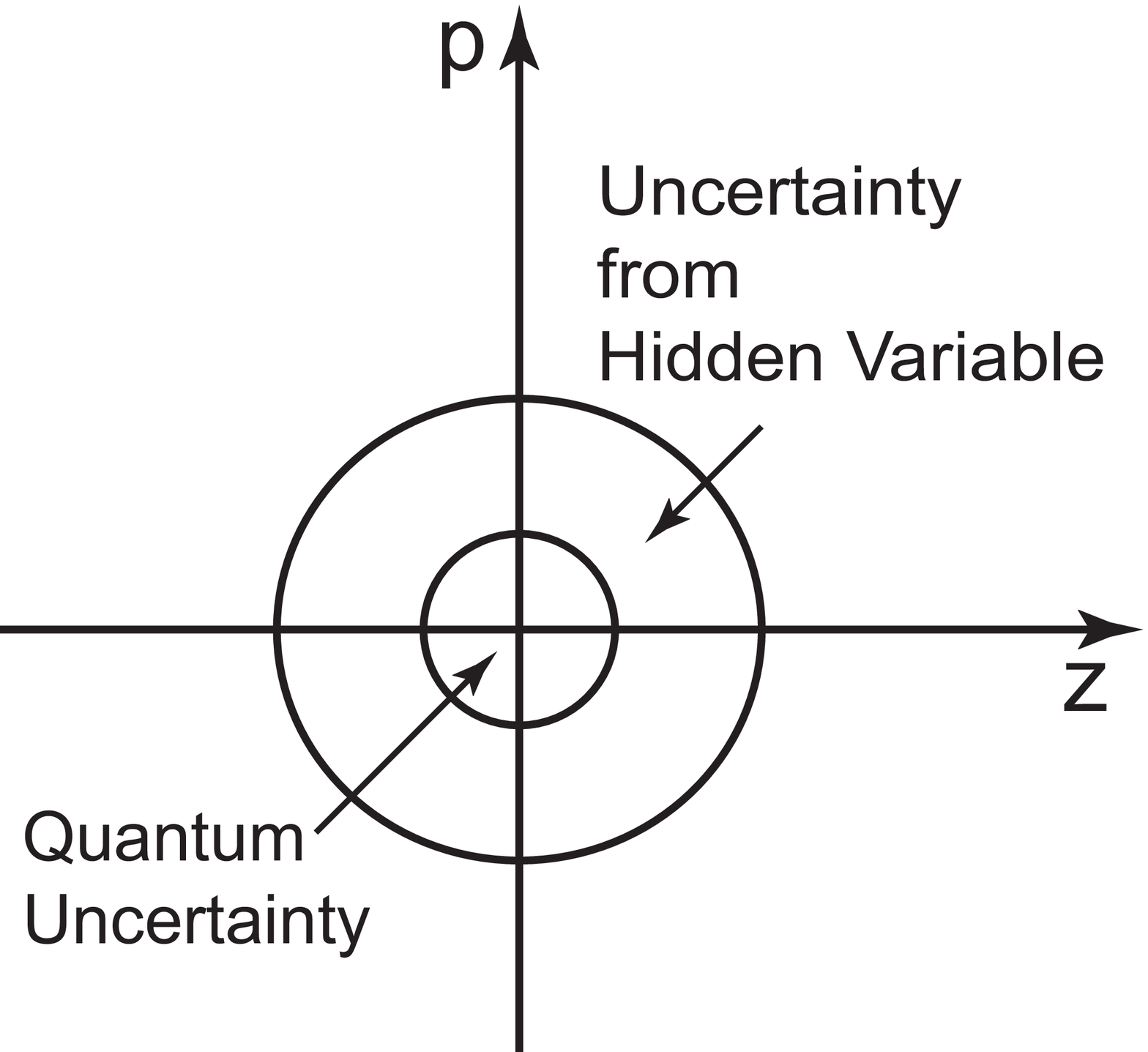}  }
\caption{Probability distribution of the two-oscillator system, which can
 also be used for the covariant harmonic oscillators and the two-photon
 coherent states.  One of the coordinates is observed and the other is
 is hidden in Feynman's rest of the universe.   In the phase-space picture
 of quantum mechanics, the small circle indicates the minimal uncertainty
 when the hadron is at rest.  The statistical uncertainty is added when
 the hadron moves.  This is illustrated by a larger circle.   The radius
 of this circle increases by $\sqrt{\cosh(2\eta)}$ as the hadron  picks
 up the speed while the time-separation varible remains as a hidden
 variable.}\label{restof33}
\end{figure}

Since we are not measuring the time-separation variable, we have
to take the trace of the matrix with respect to the t variable.
The resulting density matrix is
\begin{equation}\label{den33}
 \rho(z,z') = \left(\frac{1}{\pi \cosh(2\eta)}\right)^{1/2}
 \exp{\left\{-\frac{1}{4}\left[\frac{(z + z')^2}{\cosh(2\eta)}
                   + (z - z')^2\cosh(2\eta)\right]\right\}} ,
\end{equation}
The standard way to measure this ignorance is to calculate the
entropy defined as
\begin{equation}
         S = - Tr\left(\rho \ln(\rho)\right)  ,
\end{equation}
which, for density matrix of Eq.(\ref{den33}), becomes
\begin{equation}
  S = (\cosh^2\eta) \ln(\cosh^2\eta)
             - (\sinh^2\eta) \ln(\sinh^2\eta).
\end{equation}
The quark distribution $\rho(z,z)$ becomes
\begin{equation}\label{den55}
 \rho(z,z) = \left(\frac{1}{\pi \cosh(2\eta)}\right)^{1/2}
 \exp{\left(\frac{-z^2}{\cosh(2\eta)}\right) }.
\end{equation}
The width of the distribution becomes $\sqrt{\cosh\eta}$, and
becomes wide-spread as the hadronic speed increases.  Likewise,
the momentum distribution becomes wide-spread~\cite{knp86}.
This effect can be seen from the Wigner phase-space distribution
function defined as
\begin{equation}
W(z,p) = \int \rho(z + y, z - y) e^{2ipy} dy .
\end{equation}
For the density matrix of Eq.(\ref{den55}), this Wigner function
becomes
\begin{equation}
W(z,p) = \frac{1}{\cosh(2\eta)}
\exp{\left\{ -\left(\frac{z^2 + p^2}{\cosh(2\eta)}\right) \right\} } .
\end{equation}
This position-momentum distribution is illustrated in Fig.(\ref{restof33}).
\par
If the hadron is at rest, the time-separation variable does not play
any role in the system.  The uncertainty is purely from Heisenberg's
uncertainty relation.  If the hadron moves, and if we do not observe
the time-separation variable, there is an added uncertainty as
described in Fig.~\ref{restof33}.  This is exactly what we expect
from hidden variables.  Indeed, the time-separation variable is hidden
in Feynman's rest of the universe.
\par
Let us go back to Eq.(\ref{cwf55}) of Sec.~\ref{diracqm}.  This is a
series expansion of the squeezed ground state wave function.  This
formula serves as the two-photon coherent state in quantum optics.
If we do not observe one of the photons, the mathematics is exactly
the same as the one we carried out for the Lorentz-squeezed hadron we
presented in this report.  The increase in entropy and uncertainty in
this case has been discussed in the literature~\cite{ekn89}.  This
example allows us to study the hidden time-separation variable in terms
of what we observe in the real world.

\section*{Concluding Remarks}
In Einstein's Lorentz-covariant world, the time separation exists
whenever there is a space separation like the Bohr radius.  However,
this variable is never mentioned in the Copenhagen interpretation of
quantum mechanics.
\par
In order to see what happens if this variable is included, we
started with Feynman's phenomenological equation for the quarks inside
the moving hadron.  It is shown that there is a set of solutions
possessing the symmetry of Wigner's little group dictating internal
space-time symmetry of particles in the Lorentz-covariant world.
\par
It is noted that this set of solutions constitutes the mathematical
basis for two-photon coherent states, where both photons are observable,
but we can also study the case where one of them is not observable.
\par
With these tools, we have shown that the time-separation variable is
hidden in Feynman's rest of the universe.  This causes an increase
in entropy and uncertainty, as we expect from hidden variables.
\par
At this time, we are not able to say anything about possible hidden
variables behind Heisenberg's uncertainty principle illustrated
by a small circle in the phase-space picture in Fig~\ref{restof33}.
This could be a similar problem or an entirely different problem.
We do not know.

\end{document}